\documentclass[amsmath,twocolumn,amssymb,showpacs]{revtex4}
\usepackage{amssymb}
\usepackage{mathrsfs}
\usepackage{amsfonts}
\usepackage{graphicx}
\usepackage{dcolumn}
\usepackage{amsmath}
\usepackage{epsfig}

\begin{document}

\title{Quantum Secret Sharing by applying Analytic Geometry}

\author{Ruilong Liu$^{1}$}
\email{liuruilongfuture@gmail.com}

\author{Ying Guo$^{1,2}$}
\email{sdguoying@gmail.com}

\affiliation{$^{1}$School of Information Science $\&$ Engineering,
Central South University, Changsha 410083, China\\$^{2}$State Key
Laboratory on Fiber-Optic Local Area Communication Networks and
Advanced Optical Communication system, Department of Electronic
Engineering, Shanghai Jiaotong University, Shanghai 200030, China}

\begin{abstract}

In this paper, we investigate a novel $(2,2)$-threshold scheme and
then generalize this to a $(n,n)$-threshold scheme for quantum
secret sharing (QSS) which makes use of the fundamentals of Analytic
Geometry. The dealer aptly selects GHZ states related to the
coefficients which determine straight lines on a two-dimension
plane. Then by computing each two of the lines intercept or not, we
obtain a judging matrix whose rank can be used to determine the
secret stored in entangled bits. Based on the database technology,
authorized participants access to the database to obtain the secret
information and hence the secret never appears in the channel. In
this way, the eavesdroppers fail to obtain any secret by applying
various attack strategies.


\end{abstract}

\maketitle

\section{introduction}
The novel discovery that quantum effects could prevent secret
information from being eavesdropped in an insecure channel, which is
illustrated by Wiesner \cite{Wiesner}, and then by Bennett {\it et
al.} \cite{Bennett}, attracts all scientists' eyes to the quantum
cryptography. In 1999, Hillery {\it et al.} \cite{Hillery} applied
three-particle and four-particle Greenberger-Horne-Zeilinger (GHZ)
states to implement an initial QSS scheme. From then on, quantum
secret sharing plays an increasingly significant role in the secret
protection of quantum cryptography.

Specifically, the QSS scheme is known as a threshold scheme
\cite{threshold scheme}. Suppose that we divide a message into $n$
pieces such that any $k$ of the $n$ pieces could recover the
message, but any sets of $k-1$ or fewer pieces failed to, so the
scheme is called a $(k,n)$-threshold scheme. In this way, people can
forbid some dishonest participants, at most $k-1$ in this threshold
scheme, from knowing the whole message without the help of the
honest ones.

Motivated by the unprecedented developing pace of QSS, a myriad of
protocols and schemes have been proposed. For instance, cooperated
with quantum secure direct communication (QSDC), Zhang presented a
novel concept of quantum secret sharing \cite{Zhang 2005}, which
proved to be known as QSS-SDC, then he made use of swapping quantum
entanglement of Bell states to propose a multiparty QSS protocol
which was based on the classical-message (QSS-CM) \cite{QSSCM}.
Moreover, a $(n,n)$-threshold scheme based on GHZ state and
teleportation was proposed by Wang {\it et al.} \cite{Wang 2007};
after that Han and Liu {\it et al.} \cite{Han 2008} illustrated a
multiparty QSS-SDC scheme which used photons and random phase shift
operations (RPSOs). Furthermore, a quantum secret sharing experiment
based on a single qubit protocol in telecommunication fiber was
reported by Bogdanski {\it et al.} \cite{J. B 2008}; then Markham
and Sanders \cite{Markham 2008} illustrated the graph states for
quantum secret sharing. Recently, Sarvepalli {\it et al.} \cite{S
2009} illustrated the method to share secret by using quantum
information, which converted a set of pure $[[n,1,d]]_{q}$
Calderbank-Shor-Steane (CSS) codes to perfect secret sharing
schemes; after that the first experimental generation and
characterization of a six-photon Dicke state was demonstrated by
Prevedel {\it et al.} \cite{P 2009}, then they showed the
applications in multiparty quantum networking protocols such as
quantum secret sharing, open-destination teleportation, and
telecloning.

Hence, with the recent high-speed developments of large-capacity
storage technology and its appendant database technology, it is
innovative for us to store the GHZ states and their corresponding
coefficients of straight lines in a particular table which belongs
to a specific database. Also, the security mechanisms of databases
protect the information in databases from being known by uncertified
users. Therefore, after considering recent discoveries both on
theories and experiments, we make use of these characteristics to
implement a novel QSS scheme which is based on the fundamentals of
Analytic Geometry. In fact, the GHZ states particles just work as
the carriers which carry these states to the participants, and the
secret never appears in the channel. After the GHZ measurements on
received particles, certified participants could access to the
database and search for the coefficients of straight lines. The
positions of lines are on behalf of the secrets which actually store
in entangled bits. Accordingly, the participants judge the lines'
positions together and obtain the secret.

In this paper, we introduce some elementary knowledge of this
scheme, and detailedly describe the implementation of the
$(2,2)$-threshold QSS scheme in section II. In section III, we
generalize the $(2,2)$-threshold QSS scheme to a $(n,n)$-threshold
one and depict this implementation in detail. Then in section IV, we
consider some different attack strategies and analyze the security
of the scheme by calculating the corresponding maximum Von Neumann
entropy.

\section{Two-Party quantum secret sharing by using the Principles of Analytic Geometry}

Before we illustrate the scheme, we should review some basic theory
of Analytic Geometry. As is known to all, every straight line on a
two-dimension plane has an equation of the following form:
\begin{equation}
Ax+By+C=0,
\end{equation}where A and B are not both zero. Conversely, if A and
B are not both zero, then every equation which likes the Eq.(1)
determines a straight line. Furthermore, if two straight lines don't
intersect on a two-dimension plane, they must parallel each other on
it. (In this paper, we ignore the situation that the two lines
coincide each other, which can be understood as a unique case that
they are parallel.)

We should prepare some entangled bits as the secrets being shared.
In detail, the secrets are actually stored in these entangled bits.
We randomly choose two bits which are given by Eq.(2) as the secret
``$M_{0}$" and ``$M_{1}$". For instance, we choose the base
$|S\rangle_{1}=|00\rangle-|11\rangle$ as the secret ``$M_{0}$", and
select the $|S\rangle_{3}=|01\rangle-|10\rangle$ as the secret
``$M_{1}$". And the coefficients of entangled bits are decided by
all participants before every communication. In this way, the
entangled bits work as the secret shared in the following scheme.
Considering the orthonormal bases of Bell states, we obtain the
general form of the entangled bits' bases:
\begin{equation}
|S\rangle_{i}=a|\varphi_{i}\psi_{i}\rangle+(-1)^{i}b|(1-\varphi_{i})(1-\psi_{i})\rangle,
\end{equation} where
\begin{equation}
\varphi_{i}\psi_{i}=
             \begin{cases}
              00 & i\in(0,1),\\01 & i\in(2,3).
              \end{cases}\notag
\end{equation} In fact, the four vectors create a quantum system.

Let's consider the general form of orthonormal bases of GHZ states
given as:
\begin{equation}
|GHZ\rangle_{i}=\frac{1}{\sqrt{2}}(|\varphi_{i}\psi_{i}\phi_{i}\rangle+(-1)^{i}|(1-\varphi_{i})(1-\psi_{i})(1-\phi_{i})\rangle),
\end{equation}where
\begin{equation}
\varphi_{i}\psi_{i}\phi_{i}=
             \begin{cases}
              000 & i\in(0,1),\\001 & i\in(2,3),\\ 010 & i\in(4,5),\\100 &
              i\in(6,7).
              \end{cases}\notag
\end{equation}
We can prepare three qubits in a GHZ state of the bases above.
Hence, the well-prepared particles can be used to carry GHZ states
to participants.

After these preparations talked above, we propose a novel two-party
quantum secret sharing shceme. It can be achieved by the following
four steps:

\begin{figure}
\begin{center}
\includegraphics [width=85mm,height=72mm]{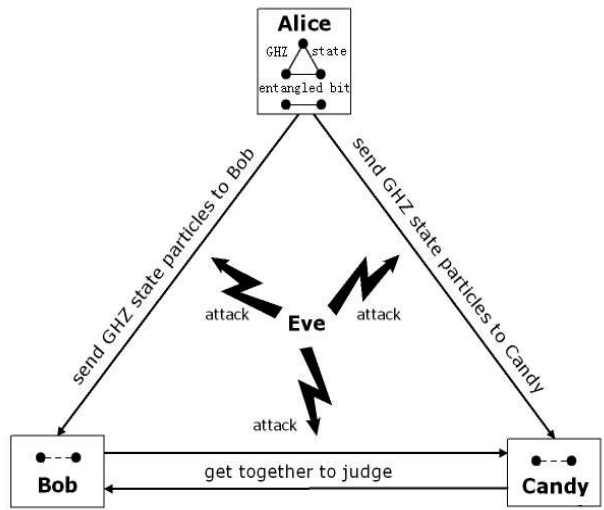}\\
\caption{The figure shows the implementation of $(2,2)$-threshold
QSS scheme based on Analytic Geometry. The triangle connects the
three qubits in GHZ state. The dashed lines which connect two qubits
represent the two participants have obtained the shared entangled
bits.}
\end{center}
\end{figure}

(i) The dealer Alice randomly chooses two entangled bits given by
Eq.(2) as the secret (e.g. ``$M_{0}$" and ``$M_{1}$") all by
herself. Then she negotiates with Bob and Candy to make sure the
coefficients of the bits, but she never tells them the bases she had
chosen. Besides, she prepared qubits in GHZ states.

In order to make the GHZ states of the particles relate to specific
entangled bits, Alice must properly selects the GHZ states.

(ii) Alice sends the GHZ state particles to Bob and Candy via a
public insecure channel, respectively. After received the triplets,
Bob and Candy implement GHZ state measurements on the received
particles.

(iii) According to result of the measurement, Bob and Candy look up
each separate table in database (e.g. Table I) to obtain a set of
coefficients of a specific straight line, then they get together to
find out whether the two straight lines intersect or not. If the two
lines parallel, the secret Alice shared is ``$M_{0}$", otherwise the
secret is ``$M_{1}$".

(iv) After the judgement, Alice declares the base of the entangled
bit she had chosen in the public channel. In this way, both Bob and
Candy get to know the secret shared by Alice.

\begin{table}
  \renewcommand{\arraystretch}{1.4}
  \centering
  \caption{Bob and Candy get a set of coefficients of the specific straight lines
  as a form $ax+by+c=0$ after the GHZ state measurement, then together find out
  whether the two lines intersect or not. To each specific set of the coefficients
  which satisfies $\frac{a}{a'}= \frac{b}{b'}$ (where $a',
  b'\neq0)$, the two straight lines are parallel each other, so the secret
  of Alice is ``$M_{0}$". If the two lines intersect at a
  single point, the secret is ``$M_{1}$".}

\begin{tabular}{|cp{5mm}p{6mm}p{5mm}p{5mm}|cp{5mm}p{6mm}p{5mm}p{5mm}|}
  \hline
  \multicolumn{5}{|c|}{\itshape Bob' table}&\multicolumn{5}{c|}{\itshape Candy's table} \\
  $|GHZ\rangle_{i}$&&$a$&$b$&$c$&$|GHZ\rangle_{j}$&&$a'$&$b'$&$c'$ \\
  \hline
  $|GHZ\rangle_{0}$ &&$a_{0}$&$b_{0}$&$c_{0}$&$|GHZ\rangle_{0}$&&$a'_{0}$&$b'_{0}$&$c'_{0}$ \\
  $|GHZ\rangle_{1}$ &&$a_{1}$&$b_{1}$&$c_{1}$&$|GHZ\rangle_{1}$&&$a'_{1}$&$b'_{1}$&$c'_{1}$ \\
  $|GHZ\rangle_{2}$ &&$a_{2}$&$b_{2}$&$c_{2}$&$|GHZ\rangle_{2}$&&$a'_{2}$&$b'_{2}$&$c'_{2}$ \\
  $|GHZ\rangle_{3}$ &&$a_{3}$&$b_{3}$&$c_{3}$&$|GHZ\rangle_{3}$&&$a'_{3}$&$b'_{3}$&$c'_{3}$ \\
  $|GHZ\rangle_{4}$ &&$a_{4}$&$b_{4}$&$c_{4}$&$|GHZ\rangle_{4}$&&$a'_{4}$&$b'_{4}$&$c'_{4}$ \\
  $|GHZ\rangle_{5}$ &&$a_{5}$&$b_{5}$&$c_{5}$&$|GHZ\rangle_{5}$&&$a'_{5}$&$b'_{5}$&$c'_{5}$ \\
  $|GHZ\rangle_{6}$ &&$a_{6}$&$b_{6}$&$c_{6}$&$|GHZ\rangle_{6}$&&$a'_{6}$&$b'_{6}$&$c'_{6}$ \\
  $|GHZ\rangle_{7}$ &&$a_{7}$&$b_{7}$&$c_{7}$&$|GHZ\rangle_{7}$&&$a'_{7}$&$b'_{7}$&$c'_{7}$ \\
  \hline
\end{tabular}

\end{table}

In fact, Alice should have known the table which Bob and Candy keep
in advance. Therefore, in step (i) she can decide which GHZ state
bases to choose to prepare the triplets aptly. In Table I, for
example, we intend to make the coefficients satisfy the following:
\begin{equation}
\frac{a}{a'} = \frac{b}{b'}
\end{equation}
or
\begin{equation}
\frac{a}{a'}\neq \frac{b}{b'}.
\end{equation}Thus if Alice wants to share the secret ``$M_{0}$", she
randomly chooses GHZ states from the table which satisfy Eq.(4);
otherwise, she will use the states shown as Eq.(5). Therefore, after
the separate measurements, Bob and Candy will find out whether their
straight lines parallel or intersect when they get together.

\section{Extension of two-party quantum secret sharing}
Suppose that N parties would like to take part in this secret
sharing, we should generalize the above two-party QSS scheme to a
N-party one. If each two of the N lines are parallel each other, the
secret is ``$M_{0}$". On the other hand, if each two are
intersecting, the secret proves to be ``$M_{1}$". Similar to the
proposed two-party QSS scheme, we get the following four steps:

(i) The dealer Alice randomly chooses two entangled bits shown as
Eq.(2) all by herself. Then she negotiates with all other
participants to make sure the coefficients of the bits without
telling them the bases she had chosen. What's more, she prepares the
particles by properly choosing the bases given by Eq.(3), which are
related to the corresponding coefficients.

(ii) Alice sends the well-prepared GHZ state particles to everyone
of the N parties via a public insecure channel. After received the
triplets, every member implements a GHZ state measurement on the
accepted particle respectively.

(iii) According to result of the measurements, everyone, such as
Bob, looks up his individual table in database (e.g. Table I) to
obtain a set of coefficients of a specific straight line. Then N
members get together to judge whether each two lines are
intersecting or paralleling. After that Alice declares the bases of
entangled bit she had chosen via the public channel.

\begin{figure}
\begin{center}
\includegraphics [width=30mm,height=30mm]{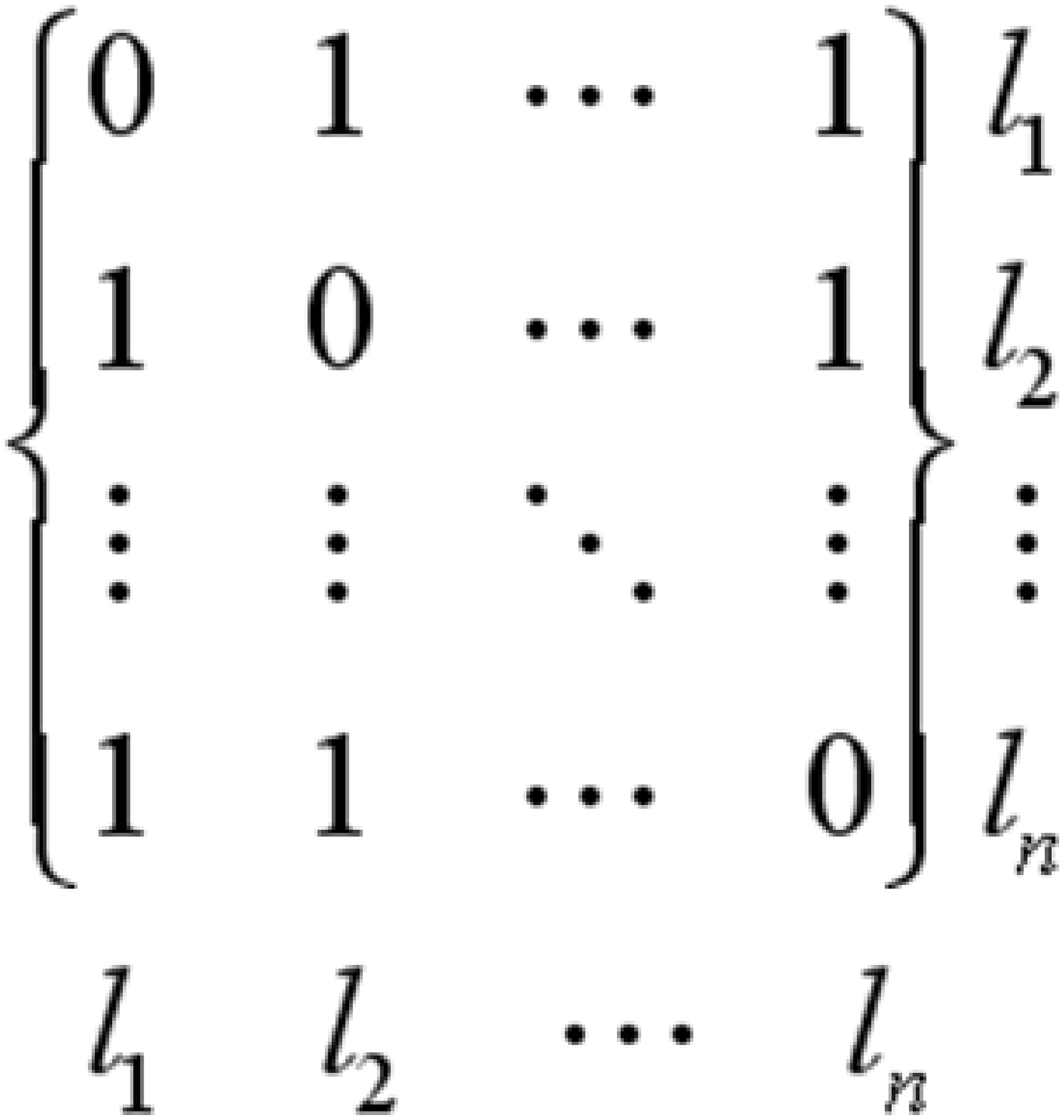}\\
\caption{This is a judgement matrix used to judge whether each two
of the N lines are paralleling or intersecting. If $l_{i}$ and
$l_{j}$ ($l_{i}$ represents the label in the row, and $l_{j}$ means
the label in the column) parallel each other, the number is ``0"; if
$l_{i}$ and $l_{j}$ intersect, the number at corresponding position
is ``1". Attention, we regulates the numbers in the diagonal line
are all ``0". }
\end{center}
\end{figure}

In order to judge whether each two of the N lines are parallel or
intersecting, we first number the N lines as $l_{1},l_{2},\ldots,
l_{n}$. Then we create a matrix shown as FIG. 2, the ``0" in the
matrix means the $l_{i}$ and $l_{j}$ ($l_{i}$ represents the label
in the row, and $l_{j}$ means the label in the column) are
paralleling, the ``1" means the $l_{i}$ and $l_{j}$ are
intersecting. The numbers in the diagonal line of the matrix are all
``0".

(iv) Participants obtain the secret shared by Alice by judging the
rank of the $n$ order square matrix: if the rank is zero, which
means the matrix is a zero matrix, the secret is ``$M_{0}$"; if the
rank of the matrix is $n$, the secret is ``$M_{1}$"; otherwise, the
results contain no information.

\section{security analysis}
As is known to all, we are always challenged by eavesdropping, for
the channels used for sending the particles are insecure. What's
worse, one dishonest participant or more may cheat the others to
obtain the secret. Hence, in order to analyze the security of the
illustrated scheme, we introduce Von Neumann entropy to evaluate the
maximum information entropy eavesdroppers can get.

Similar to the Shannon entropy which measures the uncertainty with a
classical probability distribution, Von Neumann entropy is described
with density operators replacing probability distributions. Hence,
we have the density operator \cite{density operator}:
\begin{equation}
\rho=\sum_{i} p_{i}|\phi_{i}\rangle\langle\phi_{i}|.
\end{equation}In the above equation, $|\phi_{i}\rangle$ proves to be one
of various states in a quantum system, and $p_{i}$ is the
corresponding probability. Hence Von Neumann defined the entropy of
a quantum state by the following equation:
\begin{equation}
S(\rho)\equiv-tr(\rho\log\rho).
\end{equation} In this formula logarithms are taken to base two.
If $\lambda_{x}$ are the eigenvalues of $\rho$, then Von Neumann's
definition can be re-expressed as:
\begin{equation}
S(\rho)=-\sum_{x}\lambda_{x}\log\lambda_{x}.
\end{equation}

After the introduction, let's consider the eavesdropping strategies
and evaluate the security by computing the maximum Von Neumann
entropy.

\subsection{Eavesdropping In the Insecure Channel}
If the eavesdropper Eve wants to obtain the information from the
channel, he has two methods: (i)Eve intercepts and captures the
particles from the sequences which Alice sends to the participants;
(ii) Eve intercepts the particles from the channel, then uses the
GHZ measurements to measure the particles and resends them to the
corresponding parties respectively.

If Eve applies the (i) method to eavesdrop the channel, then N
parties fail to realize that the eavesdroppers have intercepted the
particles. In order to obtain the secret, Eve should measure the
particles he has captured. Before that, he must access to the
separate tables in database which record the coefficients of
straight lines from different parties. Otherwise it's impossible for
him to find out the coefficients. However, even though he gets all
the tables kept by the N-party, he may not get the right
coefficients selected by Alice. Eve randomly chooses a basis to
measure the captured particles, then for each participant, he has a
probability of $\frac{1}{8}$ to get the right GHZ state. Thus the
eavesdropper only gets a probability of $(\frac{1}{8})^{N}$ to
acquire all the correct coefficients of the corresponding straight
lines. In this way, Eve can hardly obtain the correct judging matrix
if $N$ is large enough. Admittedly,Eve seems to have a rather small
probability to obtain the secret. However, since he don't know the
accurate coefficients of the entangled bits, even he get to know the
bases of the bits, the eavesdropping would be meaningless.

Then let's calculate the maximum Von Neumann entropy eavesdroppers
can get. According to column matrix form of the four vectors which
are obtained from Eq.(2), we get the outer product operator
$|\phi_{i}\rangle\langle\phi_{i}|$. Since the probability $p_{i}$ of
each outer product operator is $(\frac{1}{8})^{N}$, we obtain the
density operator:
\begin{eqnarray}
  \rho &=& diag\left(
                 \begin{array}{cccc}
                 \frac{a^2}{ 2^{3N-1}}, & \frac{a^2}{ 2^{3N-1}}, & \frac{b^2}{
                 2^{3N-1}}, & \frac{b^2}{ 2^{3N-1}}
                 \end{array}
               \right)
  .
\end{eqnarray}Therefore, based on the fundamentals of matrix, we
obtain the eigenvalues by calculating the $|\rho-\lambda I|=0$,
where $I$ is a unit matrix.

Moreover, since $a$ and $b$ are the coefficients of basis vectors in
the quantum system, we have $a^{2}+b^{2}=1$. According to Eq.(8), we
get the equation of maximum Von Neumann entropy:
\begin{equation}
S(\rho)=\frac{3N-1}{2^{3N-2}}- \frac{1}{2^{3N-3}}(a^{2}\log a+
b^{2}\log b).
\end{equation} Suppose that $N=2$, $a=b=\frac{1}{\sqrt{2}}$,
hence we get the maximum Von Neumann entropy:
\begin{eqnarray}
S(\rho)&=&\frac{3}{8} (bit).
\end{eqnarray} In other words, by applying this attack strategy
the maximum Von Neumann entropy Eve can get is only $\frac{3}{8}$
bit.

In order to avoid this attack strategy, all participants of this
communication, including Alice, regulates the length of these
sequences in advance. Or Alice just declares the length of the
series before she sends these sequences. If the number of particles
fails to equal to the declared amount, the corresponding
participants inform Alice that eavesdroppers do exist in the
channel, then they abort this communication and replace the current
channel with another one.

The (ii) attack strategy applied by Eve will surely be detected, for
the GHZ measurements will destroy the original state of the
particles. After measurements on these received particles, N parties
get together to judge if each two are intersecting or paralleling
and create the judgement matrix. If the rank of the matrix is
neither $0$ nor $N$, N participants know that at least one of them
has been eavesdropped, hence they terminate the current
communication and seek for another channel. Eve only has a
probability of $\frac{1}{8}$ to obtain the correct GHZ state of a
straight line, and the probability to get all the coefficients is
$(\frac{1}{8})^{N}$, which approaches to 0 if N is large enough.
Furthermore, even though he gets all the correct GHZ state shared by
Alice, he fails to know the coefficients of the entangled bits, he
would never get any information in this way. Suppose that $N=2$ and
$a=b=\frac{1}{\sqrt{2}}$, the maximum Von Neumann entropy
eavesdropped by Eve also is $\frac{3}{8}$ bit.

Hence, in face of the eavesdropping in the insecure channel, this
scheme will surely avoid giving out any secret shared by Alice,
hence it is secure in any means.

\subsection{Dishonest Participants}
Besides the eavesdropping in the insecure channel, dishonest
participants may also reveal the secret. According to the scheme,
all participants decide the coefficients of the entangled bits
without knowing the bases Alice had chosen. Hence before they
together to judge, the dishonest participants could guess the
specific entangled bit from a pair, thus each shares a probability
of $\frac{1}{2}$. And a pair of entangled bits is randomly selected
from the four entangled bits given by Eq.(2). Hence in the $6$
pairs, each specific entangled bit contains a probability of
$\frac{1}{2}$ to be chosen. Accordingly, the final probability to
guess the correct basis of entangled bits equals to $\frac{1}{4}$.

Then let's compute the maximum Von Neumann entropy one dishonest
participant can get. According to the four vectors of the quantum
system, and the probability $p_{i}$ is $\frac{1}{4}$, we get the
density operator:
\begin{eqnarray}
  \rho &=& diag
\left(
  \begin{array}{cccc}
    \frac{a^{2}}{2}, & \frac{a^{2}}{2}, & \frac{b^{2}}{2}, & \frac{b^{2}}{2} \\
  \end{array}
\right).
\end{eqnarray}

We calculate the $|\rho-\lambda I|=0$ to get the eigenvalues. Hence,
we obtain the equation of maximum Von Neumann entropy as follow:
\begin{equation}
S(\rho)=1- 2(a^{2}\log a+ b^{2}\log b).
\end{equation} Suppose that $a=b=\frac{1}{\sqrt{2}}$,
then
\begin{eqnarray}
S(\rho)&=&2 (bit).
\end{eqnarray} Hence, in this case, the maximum Von Neumann entropy one dishonest participant can get is $2$ bit.

Therefore, it would be difficult for the dishonest participants to
obtain just a few pieces of the secret information. In this way,
that would be meaningless to apply this eavesdropping strategy.

\section{Conclusions}
In this paper, we have proposed a novel quantum secret sharing
scheme, which makes use of the fundamentals of Analytic Geometry as
well as matrix. In fact, we realize this scheme in a completely
different method. The GHZ state particles are just worked as
carriers of the coefficients of specific straight lines. Hence the
secret doesn't cling to the triplets which are transmitted in the
insecure channel. The dealer properly selects the entangled bits and
controls corresponding GHZ states. Then the secrets are decided by
the positions of specific straight lines which are actually
determined by the GHZ states. Since the secrets never appear in the
channel, the eavesdroppers fail to obtain any secret by intercepting
the particles transmitted in the channel.

\section*{Acknowledgements}

This work has been supported by the National Fundamental Research
Program (Grant Nos.2006CB0L0106), the National Natural Science
Foundation of China (Grant Nos.60773013, 60902044).

\end{document}